\documentstyle[galley, rotate]{mn}
\onecolumn
\input{epsf}
\def\n{{\noindent}}

\def\inc{{\int_0^{\chi_s}}}

\title[Weak Lensing Statistics] 
{Statistics of Weak Lensing at Small Angular Scales: \\
Analytical Predictions for Lower Order Moments
}

\author[Munshi \& Jain]{Dipak Munshi$^{1}$
 and Bhuvnesh Jain$^2$\\ 
$^1$Max-Planck-Institut fur Astrophysik,
Karl-Schwarzschild-Str.1, D-85740, Garching, Germany\\
$^2$Johns Hopkins University, Department of Physics,
Baltimore, MD 21218, USA}

\pagerange{000,000}

\begin{document}

\maketitle

\begin{abstract}
Weak lensing surveys are expected to provide direct measurements of the 
statistics of the projected dark matter distribution. Most analytical 
studies of weak lensing statistics have been limited to quasilinear 
scales as they relied on perturbative calculations. On the other hand, 
observational surveys are likely to probe angular scales less than 10 
arcminutes, for which the relevant physical length scales are in the 
nonlinear regime of gravitational clustering. We use the hierarchical 
ansatz to compute the multi-point statistics of the weak lensing 
convergence for 
these small smoothing angles. We predict the multi-point cumulants and 
cumulant correlators up to fourth order and compare our results with 
high resolution ray tracing simulations. Averaging over a large number of 
simulation realizations for four different cosmological models, we find 
close agreement with the analytical calculations. In combination with 
our work on the probability distribution function, these results provide
accurate analytical models for the full range of weak lensing
statistics. The models allow for a detailed exploration of
cosmological parameter space and of the dependence on angular scale
and the redshift distribution of source galaxies. 
We compute the dependence of the higher moments of the convergence 
on the parameters $\Omega$ and $\Lambda$ and on the nature of 
gravitational clustering. 
\end{abstract}

\begin{keywords}
Cosmology: theory -- large-scale structure
of the Universe -- Methods: analytical
\end{keywords}

\section{Introduction}

Weak distortions in the images of high redshift galaxies due to
gravitational lensing provide us with valuable information about the
mass distribution in the universe. The study of such
distortions provides us a unique way to probe the statistical 
properties of the intervening large-scale structure. 
Traditionally, the study of
gravitational clustering in the quasi-linear and non-linear regimes
has been done by analyzing galaxy catalogs. However such studies can only
provide us with information on how the galaxies are clustered. 
To infer the statistics of the underlying mass distribution from 
galaxy catalogs one needs a prescription 
for how the galaxies are biased relative to the dark matter. Weak 
lensing studies have the advantage that they
can directly probe the statistics of the underlying mass distribution.

Pioneering  work in this direction was done by Blandford et
al. (1991), Miralda-Escude (1991) and Kaiser (1992) based on the 
early work of Gunn (1967).  Current progress in weak lensing
can broadly be divided into two distinct categories. Whereas Villumsen
(1996), Stebbins (1996), Bernardeau et al. (1997), Jain \& Seljak
(1997) and Kaiser (1998) 
have focussed on the linear and quasi-linear regime by assuming a 
large smoothing angle, several authors have developed a numerical 
technique to simulate weak lensing catalogs. Numerical
simulations of weak lensing typically employ N-body simulations, 
through which ray tracing experiments are conducted 
(Schneider \& Weiss 1988; Jarosszn'ski et al. 1990; Lee \& Paczyn'ski
1990; Jarosszn'ski 1991; Babul \& Lee 1991;  Bartelmann \& Schneider
1991, Blandford et al. 1991). Building on the earlier work of Wambsganss
et al. (1995, 1997, 1998) detailed numerical study of lensing 
was done by  Wambsganss, Cen \& Ostriker (1998). Other recent studies
using ray tracing experiments have been conducted by Premadi, Martel
\& Matzner (1998), van Waerbeke, Bernardeau \& Mellier (1998),
Bartelmann et al (1998), Couchman, Barber \& Thomas (1998) and
White \& Hu (1999). 
While  a peturbative analysis can provide valuable information at large
smoothing angles, it can not be used to study lensing 
at small angular scales as the whole perturbative series starts to diverge.

A complete analysis of weak lensing statistics at small angular scales
is not available at present, as we do not have a corresponding analysis
for the underlying dark matter distribution. However there are several 
non-linear {\em ansatze} which predicts a tree hierarchy for 
the matter correlation functions and are thought to be successful 
to some degree in modeling results from numerical simulations. Most 
of these {\em ansatze}
assume a tree hierarchy for higher order correlation functions but 
differ in the way they assign weights to trees of the same order with 
different topologies (Balian \& Schaeffer 1989, Bernardeau \& 
Schaeffer 1992; Szapudi \& Szalay 1993). The evolution of the two-point 
correlation functions in all such approximations 
is taken as given. Recent studies by several authors 
(Hamilton et al 1991; Peacock \& Dodds 1994; Nityanada \& Padmanabhan 1994;   
Jain, Mo \& White 1995; Peacock \& Dodds 1996) have provided fitting 
formulae for the evolution of the two-point correlation function,
which can be used in
combination with the hierarchical {\em ansatze} to predict the
clustering properties of the dark matter distribution in the universe.

Cumulants and cumulant correlators are one and two point averages of
higher order correlation functions. Scocciomarro \&
Frieman (1998) have shown that it is 
possible to predict the normalized non-linear one point cumumlants or
$S_N$ parameters in the highly non-linear regime by considering co-linear
configurations of wave vectors in the perturbative expressions for 
the multi-spectra,
a method labeled hyper-extended perturbation theory (HEPT). 
Munshi et al (1999c) combined this method with Bernardeau \&
Schaeffer's (1992) ansatz to predict amplitudes associated with distinct
topological tree configurations of arbitrary order. They used this method to
compute cumulant correlators in projected galaxy
catalogs and found very good agreement. A similar analysis was done
for the case of extended
perturbation theory as proposed by Colombi et al. (1996) and was found
to be in good agreement with that of HEPT. While such studies are 
generally done for projected catalogs it was shown by Hui (1999)
that the hierarchal ansatze can also be used  to predict the skewness of 
the convergence field associated with weak lensing surveys. 

In this paper we extend earlier studies of the lower order 
cumulants of the convergence field $\kappa(\theta_0)$. 
We present a detailed theoretical analysis of cumulant correlators
in the context of weak-lensing surveys (Munshi \& Coles 1999). 
Using several realizations of weak lensing simulations 
for four different cosmological scenarios we compare the analytical results 
with numerical results for a range of smoothing angles and source redshifts.

Several approximations are used in the analytical calculations of 
the higher order cumulants of the weak lensing convergence field 
$\kappa(\theta_0)$. 
One such approximation is the Born approximation
which neglects all higher order terms in the photon propagation equation.
This means that the relation connecting the line of sight 
integration of the density field
$\delta$ with the convergence field $\kappa$ is only approximate.
Perturbative calculations were used earlier to show that such 
correction terms are negligible for lower order cumulnats (Bernardea
et al 1997). The effects of 
source clustering on the statistics of the convergence field
have been estimated by Bernardeau (1997) using perturbative
calculations. Such analyses 
are not possible in the highly non-linear regime as the entire
perturbative series starts to diverge. The only way to check the
approximations is to compare the results with ray tracing experiments. 

In section Section 2 we briefly describe the ray tracing simulations
used to check the analytical predictions. 
Section 3 presents most of the analytical 
results necessary for computing cumulants and cumulant
correlators in the highly nonlinear regime. The comparison of the
analytical and simulation results is made in 
Section 4. In section 5 we discuss
our result in a general cosmological framework. The ray tracing
experiments placed the source at unit redshift; in the 
Appendix we study how our results depend on the source redshift.

\section{The Generation of Convergence Maps from N-body Simulations}

Convergence maps are generated by solving the geodesic equations for
the propagation of light rays through N-body simulations of dark matter
clustering. The simulations used for our study are adaptive 
$P^3M$ simulations with
$256^3$ particles and were carried out using codes kindly made
available by the VIRGO consortium. These simulations can resolve
structures larger than $30h^{-1}kpc$ at $z = 0$ accurately. These
simulations were carried out using 128 or 256 processors on CRAY T3D
machines at Edinburgh Parallel Computer Center and at the Garching
Computer Center of the Max-Planck Society. 
These simulations were intended primarily for
studies of the formation and clustering of galaxies 
(Kauffmann et al 1999a, 1999b;
Diaferio et al 1999) but were made available by these authors and by the Virgo 
Consortium for this and earlier studies of gravitational lensing
(Jain, Seljak \& White 1999; Reblinski et al 1999). 

Ray tracing simulations were carried out by Jain et al. (1999) using
a multiple lens-plane calculation which implements the discrete
version of recursion relations for mapping the photon position and the
Jacobian matrix (Schneider \& Weiss 1988; Schneider, Ehler \& Falco
1992). In a typical experiment $4\times 10^6$ rays are used to trace 
the underlying mass distribution. The dark matter distribution 
between the source and the observer is projected onto $20$ - $30$ planes.
The particle positions on each plane are interpolated onto a $2048^2$ 
grid. On each plane the shear matrix is computed on this grid 
from the projected density by using Fourier space relations between the
two. The photons are propagated starting from a rectangular grid on the
first lens plane. The regular grid of photon position gets distorted 
along the line of sight. To ensure that all photons reach the
observer, the ray tracing experiments are generally done backward in time
from the observer to the source plane at redshift $z = z_s$. 
The resolution of the convergence maps is limited by 
both the resolution scale associated with numerical simulations and also 
due to the finite resolution of the grid to propagate photons. 
The outcome of these simulations are shear and 
convergence maps on a two dimensional grid. Depending on the 
background cosmology the two dimensional box 
represents a few degree scale patch on the sky. Figure 1 shows a map
of $\kappa$ for the LCDM model with a field size 3 degrees on a side and
source galaxies taken to be at $z=1$. For more details on the
generation of $\kappa$-maps, see Jain et al (1999). 

While resolution effects are important for small angular scales,
finite volume corrections play an increasingly dominant role 
for larger angular scales. Although finite volume corrections have been
extensively studied in the context of projected galaxy surveys
(Szapudi \& Colombi 1996; Munshi et al. 1999b) such results are not available 
for weak lensing surveys. The angular size of our simulation box at $z = 1$
is about $3^{\circ}$. We therefore restrict our studies to smoothing
angles less than $10'$ and use
many realizations for a given cosmological model to estimate the scatter in 
the numerical results. These realization are constructed from subsets
of the same simulation box by changing the direction of projection
for the individual lens planes (for more details on the ray tracing 
simulations see Jain et al. 1999). 

\begin{table}
\begin{center}
\caption{Cosmological parameters characterizing different models}
\label{tabsig2D}
\begin{tabular}{@{}lcccc}
&SCDM&TCDM&LCDM&OCDM\\
$\Gamma$&0.5&0.21&0.21&0.21\\
$\Omega_0$&1.0&1.0&0.3&0.3 \\
$\Lambda_0$&0.0&0.0&0.7&0.0 \\
$\sigma_8$&0.6&0.6&0.9&0.85\\
$H_0$&50&50&70&70\\
\end{tabular}
\end{center}
\end{table}

\begin{figure}
\protect\centerline{
\epsfysize = 3.8truein
\epsfbox[27 75 477 564]
{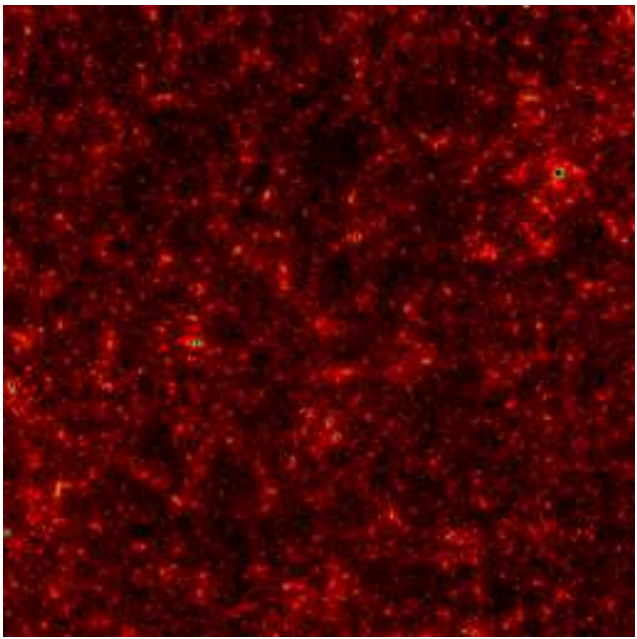} }
\caption{Convergence or $\kappa$ map generated by ray-tracing
through N-body simulations of the LCDM model. The field is 3 degrees
on a side. }
\end{figure}

\section{Cumulants and Cumulant Correlators}
\n
In this section we provide the formalism for
computing multi-point statistics using the hierarchical ansatz 
(Munshi \& Coles 1999). We use the following line element for
the background geometry:
\begin{equation}
d\tau^2 = -c^2 dt^2 + a^2(t)( d\chi^2 + r^2(\chi)d^2\Omega), 
\end{equation}
where the angular diameter distance is denoted by $r(\chi)$ and
the scale factor of the universe by $a(t)$. $r(\chi)= K^{-1/2}\sin (K^{-1/2}
\chi)$ for positive curvature, $r(\chi) = (-K)^{-1/2}\sinh
((-K)^{-1/2}\chi)$ for negative curvature and $\chi$ for zero
curvature universe. For the present value of the Hubble constant, $H_0$,
and of the mass density parameter, $\Omega_0$, we have $K= (\Omega_0 -1)H_0^2$.

\subsection{Formalism}

The statistics of the weak lensing
convergence $\kappa$ is similar to that of the projected
density filed. In what follows we will consider a small patch
of the sky where we can use the plane parallel approximation or the small
angle approximation to replace spherical harmonics by Fourier modes.
The 3D density contrast along the line of sight when projected onto
the 2D sky with the weight function $\omega(\chi)$ will provide us the
projected density contrast or the weak-lensing convergence at a
direction ${\bf \gamma}$.

\begin{equation}
\kappa({\bf \gamma}_1) = \inc {d\chi}_1
\omega(\chi_1)\delta(r(\chi){\bf \gamma}_1)
\end{equation}

\n
Assuming all the sources are at the same redshift (an
approximation often used though not difficult to modify for a more
realistic description), one can write the weight function as $\omega(\chi)
= 3/2a c^{-2}H_0^2 \Omega_m r(\chi) r(\chi_s - \chi)/ r(
\chi_s)$. Where $\chi_s$ is the comoving radial distance to the source.
Using a Fourier decomposition of $\delta$ we can write

\begin{equation}
\kappa(\gamma_1) = \inc {d\chi}_1 \omega(\chi_1) \int {d^3{\bf k} \over {(2
\pi)}^3} \exp ( i \chi_1 k_{\parallel} + r \theta k_{\perp} )\ 
\delta_{\bf k} \ ,
\end{equation}

\n
where we have used $k_{\parallel}$ and $k_{\perp}$ to denote components
of the wave vector ${\bf k}$ parallel and perpendicular to the line of
sight. 
In the small angle approximation, one assumes that $k_{\perp}$ is much
larger than $k_{\parallel}$.  $\theta$ denotes the angle between
the line of sight direction ${\bf \gamma}$ and the wave vector ${\bf k}$.

Using the definitions we have introduced above, we can compute the
projected two-point correlation function of $\kappa$ (Peebles 1980, Kaiser 1992, 
Kaiser 1998):

\begin{equation}
\langle \kappa(\gamma_1) \kappa(\gamma_2) \rangle_c = \inc d {\chi_1}
{\omega^2(\chi_1) \over r^2(\chi_1)} \int {d^2 {\bf l} \over (2
\pi)^2}~\exp ( \theta l )~ {\rm P} { \big ( {l\over r(\chi)} \big )}\ ,
\end{equation}

\n
where we have introduced ${\bf l} = r(\chi){\bf
k}_{\perp}$, a scaled wave vector projected on the
sky. The average of the two-point correlation function
smoothed over an angle $\theta_0$ with a top-hat 
window $W_2(l \theta_0)$  is given by,

\begin{equation}
\langle \kappa^2{(\gamma)} \rangle_c = \inc d {\chi_1}
{\omega^2(\chi_1) \over r^2(\chi_1)} \int {d^2 {\bf l} \over (2
\pi)^2}~ {\rm P} { \big ( {l\over r(\chi)} \big )} W_2^2(l\theta_0) .
\end{equation}

\n
Similar calculations for the volume average of the three-point correlation
function and the four-point correlation function can be expressed in
terms of integrals of the matter multi-spectrum $B_p$. The volume averages
for angular smoothing in effect smooth over a conical volume,
which in the small angle approximation is cylindrical (Munshi \& Coles 1999). The results for the three and four point function are:  

\begin{equation}
\langle \kappa^3{(\gamma)} \rangle_c = \inc d {\chi_1}
{\omega^3(\chi) \over r^6(\chi)} \int {d^2 {\bf l_1} \over (2\pi)^3}
W_2(l_1 \theta_0) \int {d^2{\bf
l_2}\over (2\pi)^2} W_2(l_2 \theta_0) \int {d^2 {\bf l_3} \over
(2\pi)^3} W_2(l_3 \theta_0) ~ {\rm B}_3 \Big ( {l_1\over r(\chi)},
{l_2\over r(\chi)},  {l_3\over r(\chi)} \Big )_{\sum {\bf l}_i = 0}
\end{equation}

\begin{eqnarray}
\langle \kappa^4{(\gamma)} \rangle_c = \inc d {\chi_1}
{\omega^4(\chi) \over r^8(\chi)} \int {d^2 {\bf l_1} \over (2\pi)^3}
W_2(l_1 \theta_0) \int {d^2{\bf
l_2}\over (2\pi)^2} W_2(l_2 \theta_0) \int {d^2 {\bf l_3} \over
(2\pi)^2} W_2(l_3 \theta_0)\int {d^2 {\bf l_4} \over 2 \pi^2} W_2(l_4
\theta_0)\\ \nonumber  ~ {\rm B}_4 \Big ( {l_1\over r(\chi)},
{l_2\over r(\chi)}, {l_3\over r(\chi)}, {l_4\over r(\chi)}  \Big )_{\sum {\bf l}_i = 0}
\end{eqnarray}

\n
In our derivation of the above results we have assumed that both the smoothing
angle $\theta_0$ and the separation angle $\theta_{12}$ are small.

Cumulant Correlatos were introduced by Szapudi \& Szalay (1997) as
normalized two-point moments of multi-point correlation functions. 
Two-point cumulant correlators 
have already been measured in projected surveys such as APM by Szapudi
\& Szalay (1997). Using the measurements they were able to separate  
contributions from different tree topologies. The concept of two-point cumulant-correlators can be further generalized to multi-point cumulant correlators.

We list below the results of our analysis for two-point cumulant
correlators for the third and fourth order in $\kappa$. The derivation of these results
is very similar to their one-point counterpart; we have
assumed that in addition to the smoothing angles being small, the
separation angle between different patches is small too. This
assumption is consistent with our assumption of the hierarchical nature
of the correlation function in the highly nonlinear. The third order 
cumulant correlator is given by,

\begin{equation}
\langle \kappa_s^2(\gamma_1) \kappa_s(\gamma_2) \rangle_c =
 \int_0^{\chi_s} { \omega^3 (\chi) \over r^4(\chi) } d \chi \int d^2
 l_1 \int d^2 \ l_2 T_3 \Big ( {{\bf l}_1 \over r (\chi)},
 {{\bf l}_2 \over r (\chi)}, {{\bf l}_3 \over r (\chi)} \Big )_{\sum
 l_i = 0} W_2(l_1
 \theta_0) W_2(l_2 \theta_0) W_2( l_3\theta_0) \exp(il_2 \theta_{12})
\end{equation}

\n 
In next section we will show that the normalized third order cumulant
correlator $C_{21}$ depends only on one hierarchal parameter $Q_3$. Its
fourth-order analogs $C_{22}$ and $C_{31}$ depend on two different
hierarchal amplitudes,
$R_a$ and $R_b$; therefore the two fourth order cumulant correlators can be
used to separate contribution from $R_a$ and $R_b$,  while their one-point 
counterpart $S_4$ can only measure an average contribution from the two 
different topologies. The two fourth order cumulant correlators are,  

\begin{eqnarray}
\langle \kappa_s^3(\gamma_1) \kappa_s(\gamma_2) \rangle_c =
\int_0^{\chi_s} { \omega^3 (\chi) \over r^4(\chi) } d \chi \int d^2
 l_1 \int d^2 l_2  \int d^2 l_3
W_2(l_1 \theta_0) W_2(l_2 \theta_0) W_2( l_3\theta_0) \exp(il_3 \theta_{12})
 \\ \nonumber
 T_3 \Big ( {{\bf l}_1 \over r (\chi)},
 {{\bf l}_2 \over r (\chi)}, {{\bf l}_3 \over r (\chi)}, {{\bf l}_4
\over r (\chi)} \Big )_{\sum
l_i = 0} ,
\end{eqnarray}

\begin{eqnarray}
\langle \kappa_s^2(\gamma_1) \kappa_s^2(\gamma_2) \rangle_c =
\int_0^{\chi_s} { \omega^3 (\chi) \over r^4(\chi) } d \chi \int d^2
 l_1 \int d^2 l_2  \int d^2 l_3
W_2(l_1 \theta_0) W_2(l_2 \theta_0) W_2( l_3\theta_0) \exp(i(l_1+l_2)\theta_{12})
 \\ \nonumber
\times\ T_3 \Big ( {{\bf l}_1 \over r (\chi)},
 {{\bf l}_2 \over r (\chi)}, {{\bf l}_3 \over r (\chi)}, {{\bf l}_4
\over r (\chi)} \Big )_{\sum
l_i = 0} .
\end{eqnarray}

\subsection{Hierarchical {\em Ansatz}}

In deriving the above expressions we have not used any specific form for
the matter correlation hierarchy. The length scales that dominate the contribution for the small angles of interest are in the highly non-linear regime. Assuming a tree model 
for the matter correlation
hierarchy in the highly non-linear regime, one can write for the most
general case (Groth \& Peebles 1977, Fry \& Peebles 1978,  Davis \&
Peebles 1983, Bernardeau \& Schaeffer 1992, Szapudi \& Szalay 1993):

\begin{equation}
\xi_N( {\bf r_1}, \dots {\bf r_N} ) = \sum_{\alpha, \rm N-trees}
Q_{N,\alpha} \sum_{\rm labellings} \prod_{\rm edges}^{(N-1)}
\xi({\bf r_i}, {\bf r_j}).
\end{equation}

It is interesting to note that a similar hierarchy 
develops in the quasi-linear regime in the limit of vanishing variance
(Bernardeau 1992), though the hierarchal amplitudes $Q_{N, \alpha}$
are shape dependent functions in that regime. In the highly nonlinear 
regime there are some indications that these functions become
independent of shape parameters, as has been proven by studies of the
lowest order parameter $Q_3 = Q$ (Sccociamarro et al. 1998).
In Fourier space such an ansatz implies that all higher-order multi-spectra 
can be written as sums of products of the matter power-spectrum.

\begin{eqnarray}
&&B_2({\bf k}_1, {\bf k}_2, {\bf k}_3)_{\sum k_i = 0} = Q ( P({\bf
k_1})P({\bf k_2}) + P({\bf k_2})P({\bf k_3})
+ P({\bf k_3})P({\bf k_1}) ) \\ \nonumber
&&B_3({\bf k}_1, {\bf k}_2, {\bf k}_3, {\bf k}_4)_{\sum k_i = 0} = R_a
P({\bf k_1})P({\bf k_1 +
k_2}) P({\bf k_1 + k_2 + k_3})  + {\rm cyc. perm.} + R_b P({\bf
k_1})P({\bf k_2})p({\bf k_3}) + 
{\rm cyc. perm.} \\ \nonumber
\end{eqnarray}

\n
Different hierarchal models differ in the way they predict the 
amplitudes of different tree topologies. Bernardeau \&
Schaeffer (1992) considered the case where amplitudes are in general 
factorizable -- at each order one has a new ``star'' amplitude 
and higher order ``snake'' and ``hybrid'' amplitudes can
be constructed from lower order ``star'' amplitudes (see Munshi et al. 1999a,b,c,
and Melott \& Coles 1999 for a detailed description). In models proposed by
Szapudi \& Szalay (1993) it is assumed that all hierarchal amplitudes of a
given order are degenerate.

We do not use any of these specific models for clustering and only
assume the hierarchal nature of the correlation functions. Galaxy surveys
have been used to study these different {\em ansatze}. Our main motivation here is to
show that weak-lensing surveys can also provide valuable information
in this direction, in addition to constraining the matter power-spectra and
background geometry of the universe. We consider the third and fourth one point moments of $\kappa$ in terms of hierarchical amplitudes: 

\begin{eqnarray}
&&\langle \kappa^3(\gamma) \rangle_c = (3Q_3){\cal C}_3[\kappa^2_{\theta_0}] 
\label {hui} \\  
&&\langle \kappa^4(\gamma) \rangle_c = (12R_a + 4
R_b){\cal C}_4[\kappa^3_{\theta_0}] \\ 
\end{eqnarray}

\noindent
Equation ({\ref {hui}}) was derived by Hui (1998). He 
showed that his result agrees well with numerical ray tracing
experiments of Jain, Seljak and White (1998).
The two-point cumulant correlators can be written as (Munshi \& Coles 1999):

\begin{eqnarray}
&&\langle \kappa_s^2(\gamma_1) \kappa_s(\gamma_2) \rangle_c = 2Q_3
{\cal C}_3[\kappa_{\theta_0} \kappa_{\theta_{12}}] + Q_3^2 {\cal C}_3[
 \kappa_{\theta_{12}}^2] \\
&& \langle \kappa_s^3(\gamma_1) \kappa_s( \gamma_2) \rangle_c = 
6 R_a {\cal C}_4[\kappa_{\theta_0}^2 \kappa_{\theta_{12}}] +
3R_b{\cal C}_4[\kappa_{\theta_0}^2 \kappa_{\theta_{12}}] +6 R_a {\cal
C}_4[\kappa_{\theta_0} \kappa_{\theta_{12}}^2] + R_b{\cal C}_4[\kappa_{\theta_{12}}^3]   \\
&& \langle \kappa_s^2(\gamma_1) \kappa_s^2(\gamma_2) \rangle_c = 4
R_b{\cal C}_4[\kappa_{\theta_0}^2 \kappa_{\theta_{12}}]+ 4
R_a{\cal C}_4[\kappa_{\theta_0} \kappa_{\theta_{12}}^2]+ 4
R_b{\cal C}_4[\kappa_{\theta_0} \kappa_{\theta_{12}}^2]+4
R_a{\cal C}_4[ \kappa_{\theta_{12}}^3]\\ 
\end{eqnarray}

\n
We have used the following notation to simplify our presentation
and have neglected the smoothing corrections which were shown to be
small by Boschan, Szapudi \& Szalay (1994):

\begin{eqnarray}
&&\kappa_{\theta_0}  = \int d^2 {\bf l} \ P \Big ( {l \over r(\chi)} \Big )
W^2(l \theta_0) \\
&&\kappa_{\theta_{12}}  =
\int d^2 {\bf l} \ P \Big ( {l \over r(\chi)} \Big ) W^2(l \theta_0)
\exp (i l \theta_{12}) \\
&&{\cal C}_t[\kappa_{\theta_0}^m \kappa_{\theta_{12}}^n] = \int_0^{\chi_s} {
\omega^{t} (\chi) \over r^{2(t-1)}(\chi) }\kappa_{\theta_0}^m \kappa_{\theta_{12}}^n   d \chi
\end{eqnarray}

The expression for $C_t[\kappa_{\theta_0}^m \kappa_{\theta_{12}}^n]$ can be computed numerically for specific
models of the background cosmology and different values of the parameters
$m$ and $n$. $F(\chi)$ denotes the various products
of $\kappa_{\theta_0}$ and $\kappa_{\theta_{12}}$ which appear in the
above expressions containing $\chi$ dependence. The values of the
hierarchal amplitudes (which in general are insensitive to the background cosmology) can be computed from numerical simulations or from 
hyper-extended perturbation theory (Scoccimarro et al. 1998,
Scoccimarro \& Frieman 1998). This will allow a detail comparison against
simulation data. 

It is customary to define the normalized cumulants and
cumulant correlator by the following equations while studying
statistics of clustering of dark matter distribution:

\begin{eqnarray}
&& S_N' = {\langle \kappa^N(\gamma) \rangle_c \over \langle
\kappa^2(\gamma) \rangle_c^{(N-1)}} \\
&& C_{pq}' = {\langle \kappa_s^p(\gamma_1) \kappa_s^q(\gamma_2) \rangle_c
\over  \langle \kappa(\gamma)^2 \rangle_c^{(N-2)} \langle   \kappa_s(\gamma_1) \kappa_s(\gamma_2) \rangle}_c
\end{eqnarray}

\n
In the context of weak-lensing surveys it is more convenient to define the normalized cumulants and cumulant correlators as: 

\begin{eqnarray}
&& S_N = {\langle \kappa^N(\gamma) \rangle_c \over \langle
\kappa^2(\gamma) \rangle_c^{N/2}} \\
&& C_{pq} = {\langle \kappa_s^p(\gamma_1) \kappa_s^q(\gamma_2) \rangle_c
\over  \langle \kappa(\gamma)^2 \rangle_c^{(N-1)/2} \langle| \kappa_s(\gamma_1) \kappa_s(\gamma_2)| \rangle_c^{1/2}}
\end{eqnarray}

\n
The $S_N$ parameter defined above for weak-lensing studies are also
denoted by $\Sigma_N$ (Valageas 1999b) in the literature.

\begin{figure}
\protect\centerline{
 \epsfysize = 3.5truein
 \epsfbox[296 166 587 453]
 {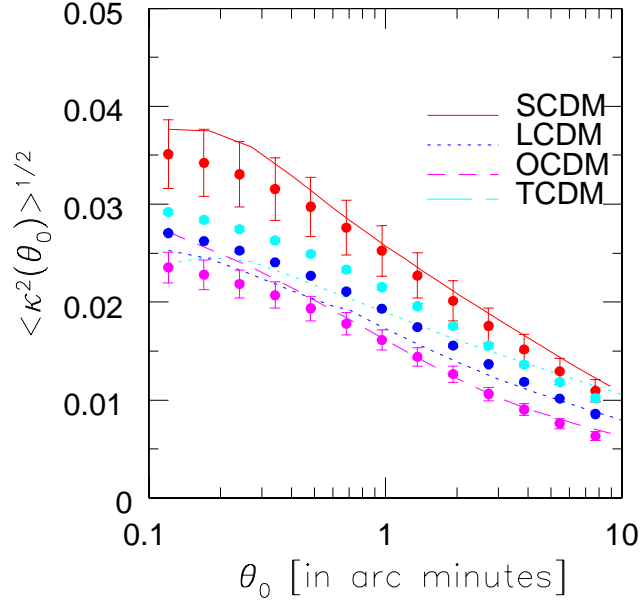} }
 \caption{Analytical predictions for the r.m.s variance of the convergence
 field $\kappa(\theta_0)$ are plotted against numerical ray tracing
 experiments as a function of the smoothing angle $\theta_0$.
Solid, dotted, short-dashed and long-dashed lines correspond to
 theoretical predictions for the SCDM, LCDM, OCDM and TCDM models. Dots
 correspond to numerical computations averaged over several different 
realizations, with the 
error bars giving the scatter between different realizations.}
\end{figure}

\begin{figure}
\protect\centerline{
 \epsfysize = 3.5truein
 \epsfbox[296 166 587 453]
 {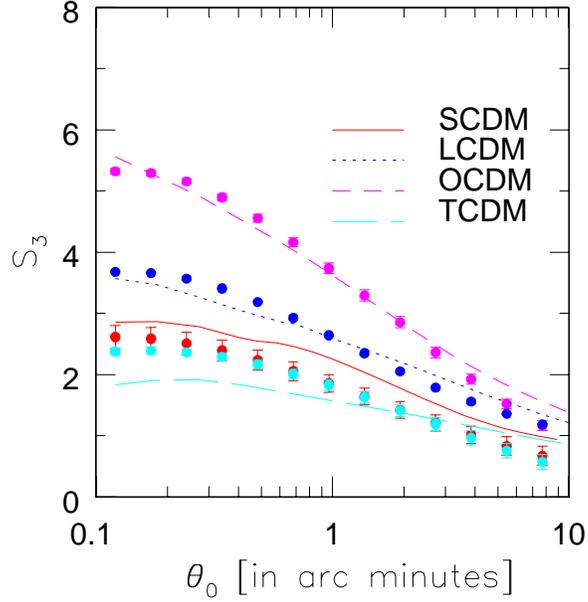} }
 \caption{Analytical predictions for the normalized skewness $S_3$ of
the smoothed  convergence
 field $\kappa(\theta_0)$. 
Solid, dotted, short-dashed and long-dashed lines correspond to
 theoretical predictions for the SCDM, LCDM, OCDM and TCDM models. Dots
 correspond to numerical computations averaged over several different 
realizations, with the error bars giving the scatter between 
different realizations. A
 hierarchical form is assumed for the 3-point correlation function of
 underlying dark matter distribution. The amplitude $Q_3$ 
is evaluated using  hyper-extended perturbation
theory (Scocciomarro \& Frieman 1998).}
\end{figure}

\begin{figure}
\protect\centerline{
 \epsfysize = 3.5truein
 \epsfbox[296 166 587 453]
 {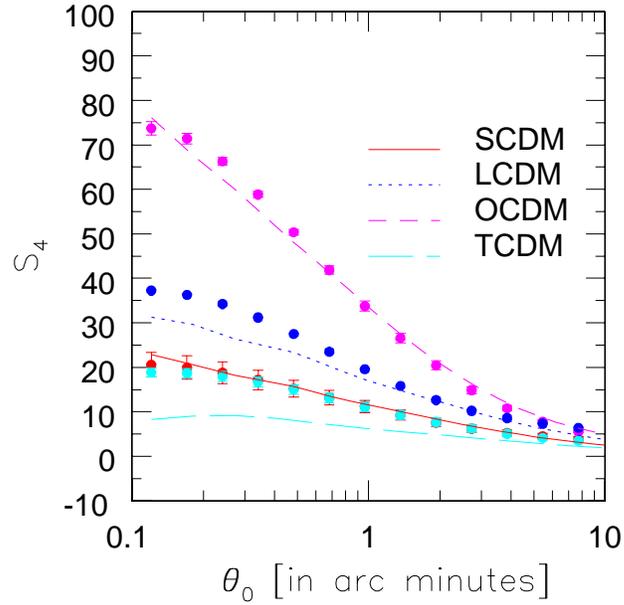} }
 \caption{Analytical predictions for the normalized kurtosis $S_4$ of
the  smoothed convergence
 field $\kappa(\theta_0)$. The lines and dots denote analytical and
numerical results as in the preceding figures. 
 The 4-point function is computed in analogy with the
three-point function shown in figure 3. }
\end{figure}

\begin{figure}
\protect\centerline{
 \epsfysize = 3.5truein
 \epsfbox[300 420 587 717]
 {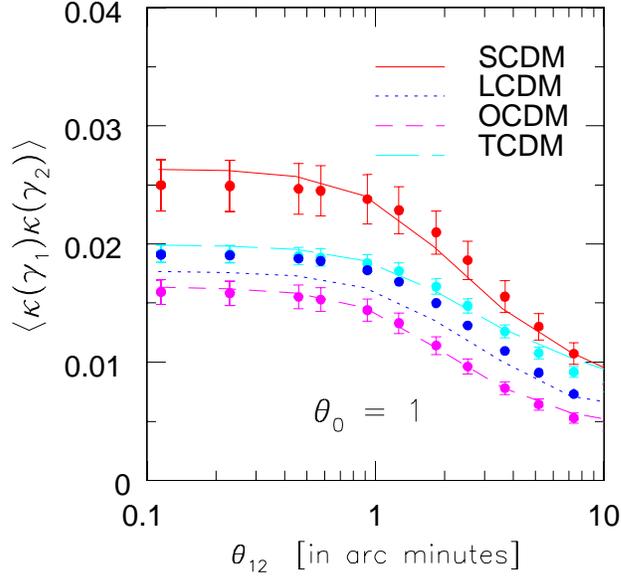} }
 \caption{Analytical predictions for the correlation function
 $\langle \kappa(\gamma_{1})\kappa(\kappa_2) \rangle$ of the smoothed 
convergence  field $\kappa(\theta_0)$ as a function of the separation angle 
$\theta_{12}$. 
The lines and dots denote analytical and
numerical results as in the preceding figures. 
The smoothing angle is 
fixed at $\theta_0=1'$. }
\end{figure}

\begin{figure}
\protect\centerline{
 \epsfysize = 3.5truein
 \epsfbox[300 420 587 717]
 {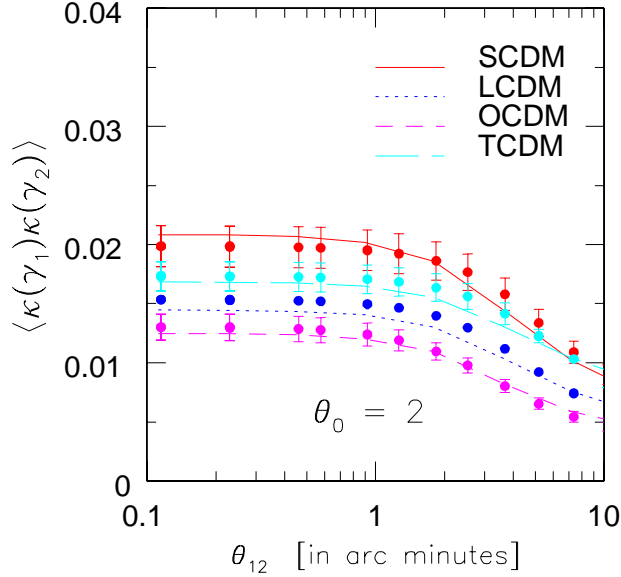} }
 \caption{The correlation function $\langle \kappa(\gamma_{1})
\kappa(\kappa_2) \rangle$ is shown as in figure 5, but with 
smoothing angle 
$\theta_0=2'$.
}
\end{figure}

\begin{figure}
\protect\centerline{
 \epsfysize = 3.5truein
 \epsfbox[300 420 587 714]
 {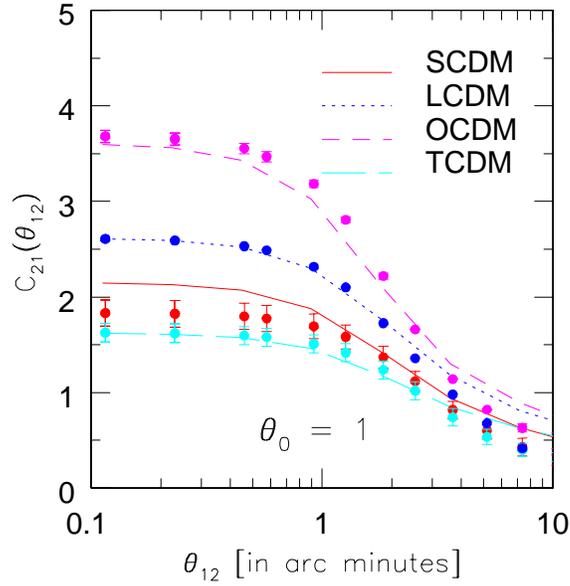} }
 \caption{The third order cumulant correlator
 $C_{21}(\theta_{12})$ of the smoothed convergence
 field $\kappa(\theta_0)$ is plotted as a function of the 
separation angle $\theta_{12}$. 
The lines and dots denote analytical and
numerical results as in the preceding figures. 
The smoothing angle is $\theta_0=1'$. 
}
\end{figure}

\begin{figure}
\protect\centerline{
 \epsfysize = 3.5truein
 \epsfbox[300 420 587 714]
 {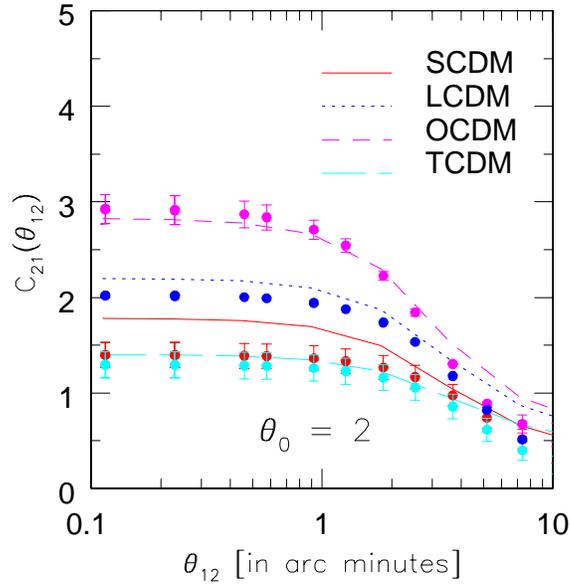} }
 \caption{The third order cumulant correlator
 $C_{21}(\theta_{12})$ is shown as in figure 7, with 
smoothing angle $\theta_0=2'$. 
}
\end{figure}

\begin{figure}
\protect\centerline{
 \epsfysize = 3.5truein
 \epsfbox[300 420 587 714]
 {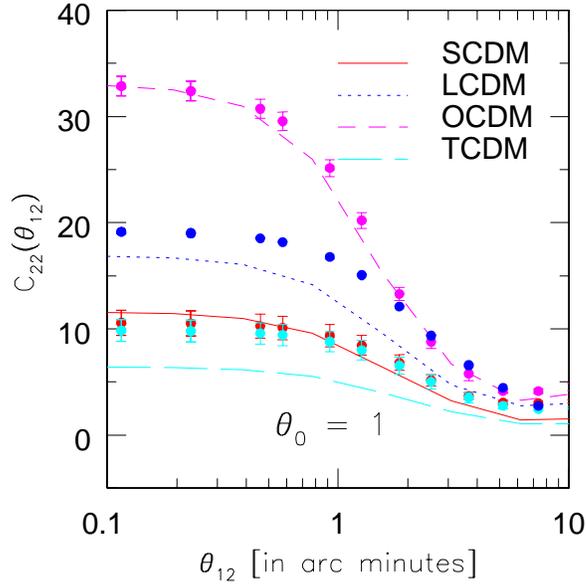} }
 \caption{The fourth order normalized cumulant
 correlator $C_{22}(\theta_{12})$ 
as a function of separation angle $\theta_{12}$.
The lines and dots denote analytical and
numerical results as in the preceding figures. 
The amplitude $Q_4$ for the four-point
 correlation function is evaluated using  hyper-extended perturbation
theory. We used Bernardeau \& Schaeffer's (1992) model for the correlation
 hierarchy to separate the topological amplitudes $R_a$ and
 $R_b$. 
}
\end{figure}

\begin{figure}
\protect\centerline{
 \epsfysize = 3.5truein
 \epsfbox[300 420 587 714]
 {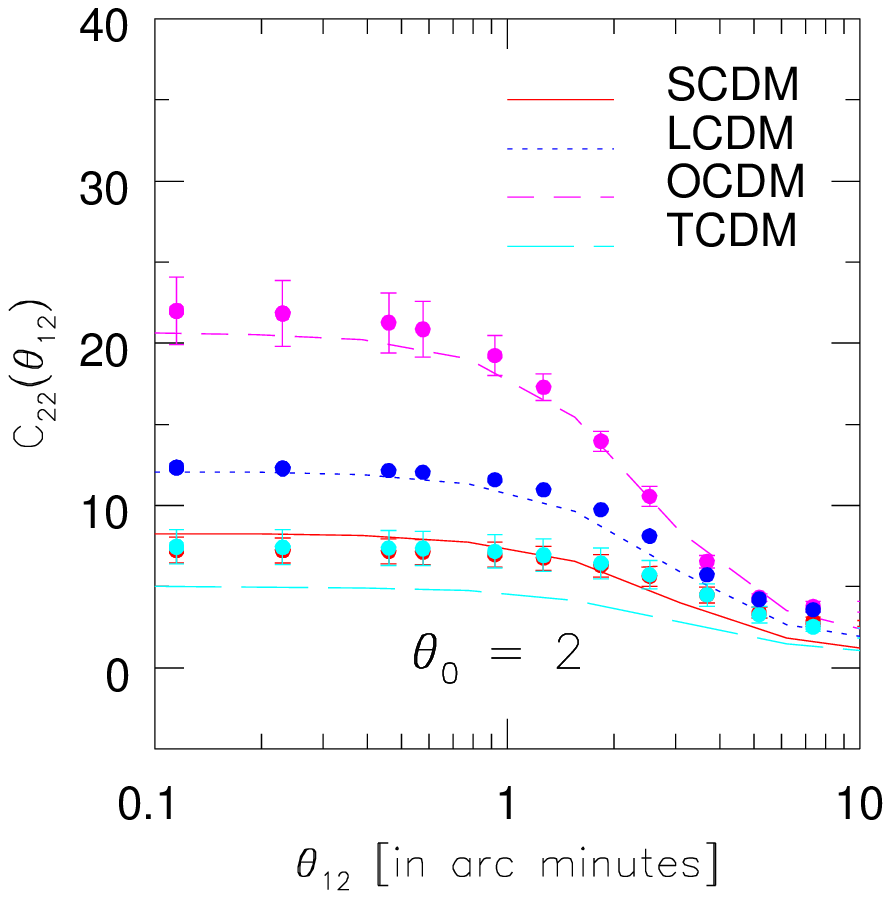} }
 \caption{The fourth order normalized cumulant
 correlator $C_{22}(\theta_{12})$ as in figure 9, with
 smoothing angle $\theta_0=2'$. 
}
\end{figure}

\begin{figure}
\protect\centerline{
 \epsfysize = 3.5truein
 \epsfbox[300 420 587 714]
 {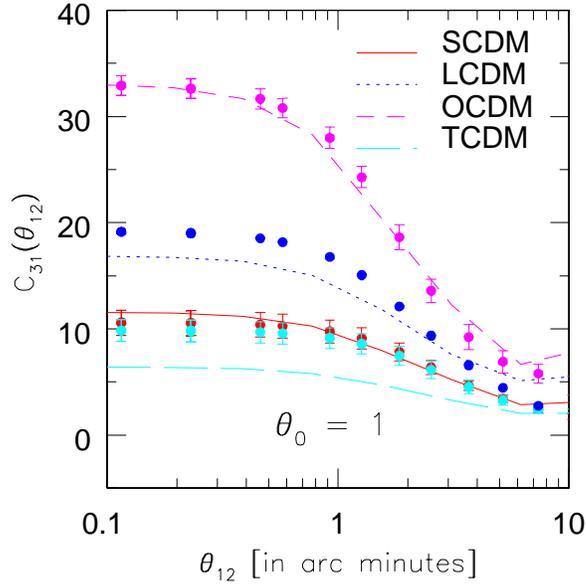} }
 \caption{The fourth order normalized cumulant
 correlator $C_{31}(\theta_{12})$ of the smoothed convergence
 field $\kappa(\theta_0)$ 
as a function of the separation angle $\theta_{12}$. 
The lines and dots denote analytical and
numerical results as in the preceding figures. 
The smoothing angle is $\theta_0=1'$. 
}
\end{figure}

\begin{figure}
\protect\centerline{
 \epsfysize = 3.5truein
 \epsfbox[300 420 587 714]
 {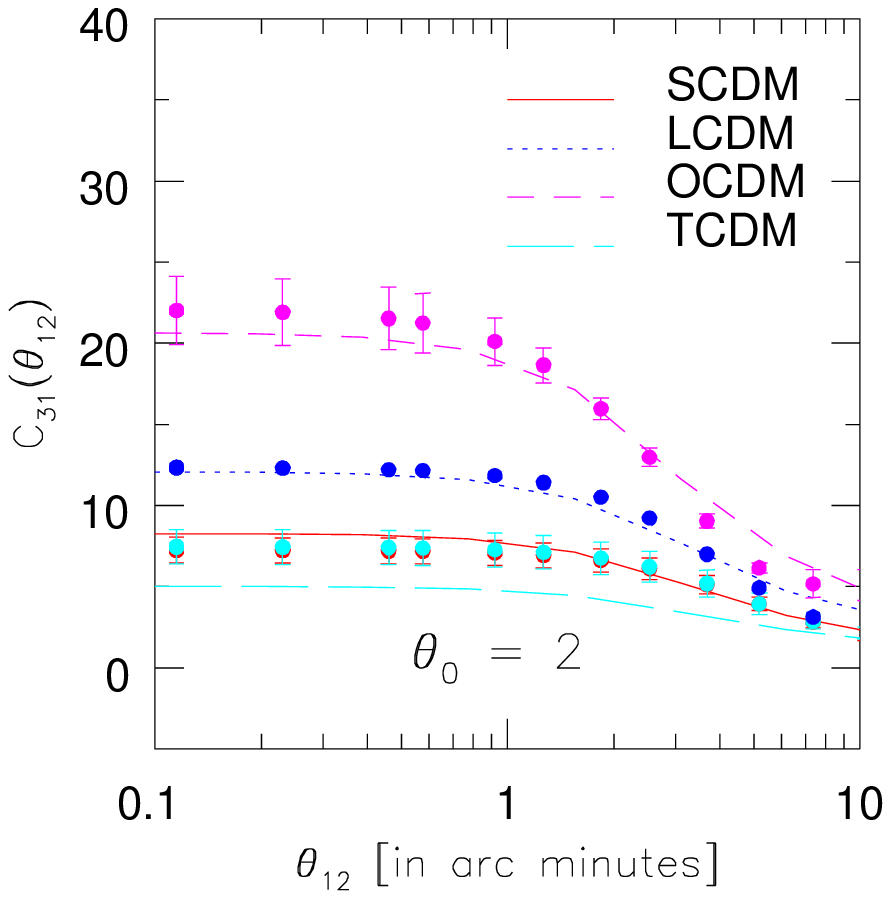} }
 \caption{
The fourth order normalized cumulant
 correlator $C_{31}(\theta_{12})$ as in figure 11, with  
smoothing angle $\theta_0=2'$.}
\end{figure}

\subsection{Modeling the non-linear power spectrum}

In numerical computations of the variance and volume averages of higher
order correlation functions, we have to include the effects of non-linearity
in the evolution of the power spectrum. The evolution of density
inhomogenieties introduces radial dependences in the expression 
for the power-spectrum. In linear theory it can be written as $P(k, {\chi})
= a^2(\chi)F^2(\chi)P(k) = D_{+}^2(\chi)
P(k)$. Where $D_{+}(\chi)$ is the linear growth factor for the evolution
of density perturbations and $F(\chi)$ depends on the background geometry 
of the universe. Lahav et al. (1991) approximated it by the
following fitting function:

\begin{eqnarray}
&&F(\chi) = {5 \over 2} \Omega_m a(\chi)^{-1} \left[ x f + {3 \over 2}
\Omega_m a(\chi)^{-1} + \Omega_K \right] ^{-1} \\
&&x = 1 + \Omega( a(\chi)^{-1} -1 ) + \Omega_{\Lambda}( a^2(\chi) -1)\ ; \ 
f = \Big ( {\Omega_m \over {a(\chi) x}} \Big )^{0.6}.
\end{eqnarray}

\n
For $a = 1$, ignoring the weak $\Omega_{\Lambda}$ dependence of the logarithmic
growth factor, the above expression gives,

\begin{equation}
F(\chi = 0) = {5 \over 2} \Omega_m ( 1 + \Omega_m^{0.6} + {1 \over 2}
\Omega_m - \Omega_{\Lambda})^{-1}
\end{equation}

On scales where the dimensionless power, $\triangle^2( k, \chi)) = 4\pi k^3 P_{\delta}( k, \chi)$, is comparable to unity
nonlinear corrections become important. This is more important for 
models with low values of $\Omega_m$. In the quasi-linear regime such
corrections can be computed using perturbation theory. 
An alternative semi-analytical approach which maps the linear power
spectrum to the non-linear spetrum was propsed by Hamilton et
al (1991) and was subsequently developed by Nityananda \& Padmanabhan
(1991); Peackok \& Dodds (1996), Jain, Mo \& White (1995) and
Padmanabhan et al. (1996)). This ansatz involves mapping the
nonlinear spectrum for a given wave number $k$ to the linear spectrum
at a wave number $k_{L}$ which are related to each other by 
$k_L = k(1 + \triangle^2( k, \chi))^{-1/3}$. The accuracy of the results
depend on the function $G$ which relates the nonlinear
power on $k$ with the linear power spectrum $k_L$
i.e. $\triangle^2(k, \chi) = G[\triangle^2_L(k_L, \chi)]$. We 
use the fitting formula of Peackock \& Dodds (1996) to describe the
non-linear power spectra as a function of length scale and redshift.
      
\subsection{Modeling hierarchical amplitudes in the non-linear
regime}

As discussed above, different models for the hierarchical amplitudes 
predict different relations between the topological amplitudes
that contribute to a given order, e.g. for the fourth order correlation functions we have two different topologies, the ``snakes'' and
``stars''. Bernardeau \& Schaeffer (1991) proposed that 
hybrid and snake topologies at a given order can be constructed 
from lower order topologies as the vertices of trees representing the 
correlation hierarchy obey a multiplicative relation. This will mean 
that $R_a = Q^2$. However note that the full prescription for $Q$, $R_a$
and other higher order ``star'' topologies  will need a more detailed
modeling  of  the non-linear density distribution. However it was
noted by Munshi et al (1999) that since at each order we have only 
one new star topologies and $S_N$ parameters at each order are 
completely determined by the topological amplitudes of the same order,
it is possible to determine these star topologies if we have a
prescription to determine $S_N$ parameters at various order. In the 
quasi-linear regime perturbation theory provides such a formalism and
Bernardeau (1992, 1994) developed a scheme to predict the $S_n$ parameters
of arbitrary order in the quasi-linear regime using a tophat smoothing.
Different extensions of perturbation theory have been suggested to predict
the one-point statistics in the highly non-linear regime. Colombi et al
(1996) proposed to replace the local spectral index $n$ in the expressions
for the $S_N$ parameters with an effective spectral index
$n_{eff}$. However it was realized that while the same $n_{eff}$ can be used
for all $S_N$ for a given power spectrum, $n_{eff}$ does not coincide
with the local slope of the non-linear power spectrum. Sccocimarro and 
Frieman (1998) on the other hand noted that the values of the $S_N$
parameters in  the co-linear
configuration in the quasi-linear regime describe the amplitudes
for these quantities in the nonlinear regime in which they become shape independent. 
We use their hyper-extended perturbation theory (HEPT) to make concrete
predictions for the $S_N$ parameters for weak lensing statistics. Their
model for the dark matter distribution predicts: 

\begin{eqnarray}
&&S_3(n) = 3\ { 4 - 2^n \over 1 + 2^{ n+ 1} } \\
&&S_4(n) = 8\ { 54 - 27~2^n + 2~3^n + 6^n \over 1 + 6~2^n + 3~3^n + 6~6^n
}.
\end{eqnarray}

\n 
We find that our results for all $S_N$ parameters at arcminute scales are very accurately
described by the local spectral index $n=-2.3$ for all
four cosmological scenarios. HEPT is very accurate in
predicting one point cumulants; it 
is possible to combine it the Benardeau \& Schaeffer (1992) ansatz to compute
the topological weights associated with different type of diagrams as well
(see Munshi et al. 1999 for more details). This is necessary to
predict the cumulant
correlators or other multi-point statistics. HEPT gives us 
$Q_3 = 2.70$ and $Q_4 = 9.52$ for $n = -2.3$, and we get  $R_a = 7.29$
and $R_b = 16.23$ for snake and star topologies, respectively.
As noted by Munshi et al (1999), one can also use EPT as proposed 
by Colombi et al. (1996) which gives us very similar results.
Finite volume N-body catalogs introduce a bias in the determination of the 
$S_N$ parameters (see.g. Colombi et al (1995)). In
particular it has been noted that the $S_N$ parameters are biased towards lower
values if determined from finite volume catalogs. Elaborate schemes were
developed to correct such finite volume effects and were tested
extensively against numerical simulations (Colombi et al. 1995;
Munshi et al. 1997). Such effects will also be important in the
determination  of various statistical quantities from ray tracing simulations, which use the outputs of N-body
simulations. We have not included finite volume
corrections in  our analysis of lower order cumulants and cumulant
correlators. A detailed study of such corrections will be presented
elsewhere.

\section{Comparing analytical predictions with numerical simulations}

\subsection{Cumulants}
Our study of one-pont statistics includes the variance, $S_3$ and $S_4$. We
have computed these quantities as a function of smoothing angle ranging
from $0.1'$ to $10'$. We have used approximately ten
realizations for each cosmological model studied by us, except for the
LCDM model where only one realization was used. Sources galaxies were taken to be at $z_s = 1$.  The use of such a large 
number of realizations gives us the opportunity to probe the degree of
fluctuation from one realization to another. We find that our results
are very accurately described by analytical approximations. Larger scales are more likely to be affected by the finite volume
corrections described above. Compared to two-point statistics, one
point cumulants are less affected by the finite size of the catalogs. In
our analysis of one-point statistics, we find very good agreement
between theory and numerical simulations, which suggests that such
effects are indeed negligible.
It is also to be noted that we have used the small angle approximation in 
our analytical computations, which is expected to hold good for the small
angular scales studied by us. 

In Figure 2 we plot the variance of the convergence field
$\kappa^2{(\theta_0)}$, smoothed with a tophat filter of smoothing angle
$\theta_0$. The variance for the SCDM model is larger than in the other models. In most of the models the analytical 
predictions show a very good agreement with numerical results. Issues related
to the effect of finite force resolution have been extensively dealt with in the study by Jain et al. (1999). We find
that for all models a smoothing angle larger than $0.25'$ is sufficient 
to remove such effects. Compared to other models, the TCDM models shows a
slight departure from the analytical results; this could be due to the
presence of large-scale power in this spectra. In Figures 3 and 
4 we have plotted the next order normalized moments,
$S_3$ and $S_4$. Again, the analytical models agree with the numerical
results very well and it is also evident that the $S_N$ parameters can directly
probe cosmological parameters, especially $\Omega_0$. At large smoothing angles one
can see that analytical results start to overestimate the $S_N$
parameters. This is related to the fact that we have considered
the hierarchical amplitudes constant with scale; in practice they go
through a transition from the non-linear regime to the quasi-linear regime.
Nonlinear estimates of the $S_N$ parameters  based on EPT or HEPT therefore over-predict such amplitudes.
Finite volume corrections also starts to become more important for
larger smoothing scales. However despite such effects  we find the 
agreement between analytical simulations and numerical results to be 
very good. A good match also indicates that the Born approximation is a 
very good approximation even for the smaller angular scales probed by us.

Throughout our studies of cumulants we have used top-hat filters for
smoothing the convergence field $\kappa$. The analytical
results we have presented here can easily be extended to other filters such 
as compensated filters, which may be more useful from an observational point
of view. We hope to present results of such analyses elsewhere.

Cumulants are normalized moments of the
one-point probability distribution function (PDF) for the convergence field.
Our analytical prediction maps cumulants of the underlying dark matter 
distribution to that of the convergence field. In a similar analysis it is
possible to map the complete PDF of the density to the convergence PDF (Munshi \& Jain 1999)

\subsection{Cumulant Correlators}  

We have measured cumulant correlators from simulations for two
different smoothing angles, $\theta_0 = 1'$ and $\theta_0 = 2'$. For each 
$\theta_0$, several separation angles
$\theta_{12}$ were considered, ranging from
from $\theta_{12} = 0.1'$ to $\theta_{12} = 10'$. We have measured
$C_{21}$ at third order and $C_{31}$ and $C_{22}$ at fourth order. 
For small separation angles when two  patches
overlap with each other, the cumulant correlators $C_{pq}$ become
equal to the moments $S_{p+q}$. The numerical results for  $C_{31}$
do correspond to $S_3$ and $C_{22}$ and
$C_{31}$ correspond to $S_4$ at small angular scales. 

In agreement
with our findings for the one-point cumulants, we find that analytical
approximations for cumulant correlators are very close to results
obtained from numerical simulations. Figures 5 and 6 show the two
point cumulant correlators, for smoothing angles $\theta_0=1', 2'$
respectively, while figures 7 and 8 show the third
order correlators. The
fourth order correlator  $C_{22}$ is shown in figures 9 and 10, with
the same smoothing angles, and $C_{31}$ is shown in figures 11 and
12. Cumulant correlators are more strongly
affected by finite volume correction compared to their one-point
counterparts. We find that for large separation angles $\theta_{12}$,
there are slight discrepancies between analytical and numerical
results. The comparison can be more appropriately made by performing
ray tracing experiments through bigger N-body boxes. 
It is also
possible that for large angular scales, the corresponding physical scales
are already in the quasilinear regime where the values of the 
hierarchical amplitudes
are smaller than their values in the highly nonlinear regime.

Since the cumulant correlators are normalized moments of the two-point joint
probability distribution function, they are directly related to
two-point statistics of the underlying fields. They can also be
related to the biasing of dark matter halos. The results of such an
analysis will be presented elsewhere.

\section{Discussion}

Ongoing weak lensing surveys with wide field CCD are likely to produce
shear maps on areas of order 10 square degrees. In the future, 
$10^{\circ} \times 10^{\circ}$ areas are also feasible, e.g. from 
the MEGACAM camera on the Canada France Hawaii Telescope and the
VLT-Survey-Telescope. Ongoing optical (SDSS; Stebbins et al. 1997) and
radio (FIRST; Kaminkowski et al. 1997) surveys can also provide useful imaging
data for weak lensing surveys. 
Such surveys will provide a very useful map of the projected
density of the universe and thus will help us test different dark
matter models and probe the background geometry of the universe. They will also
provide a unique opportunity to test different ansatze for
gravitational clustering in the highly non-linear regime in an unbiased way.
Traditionally such studies have used galaxy surveys, with the 
disadvantage that galaxies are biased tracers of the underlying mass
distribution.

Most previous studies in weak lensing statistics used a
perturbative formalism, which is applicable in the quasilinear
regime and thus requires large smoothing
angles. To reach the quasi-linear regime, survey regions must exceed 
areas of order 10 square degrees. Since existing CCD cameras typically have
diameters of $0.25^{\circ}-0.5^{\circ}$, the initial weak lensing surveys
are likely to provide us statistical information on 
small smoothing angles, of order $10'$ and less. 
This makes the use of
perturbative techniques a serious limitation,
as the relevant physical length scales  are in the highly
nonlinear regime. We have employed
a new technique based on the hierarchal ansatz to compute weak
lensing statistics on these small angular scales. By
combining the hierarchical ansatz with either hyper-extended or extended
perturbation theory, 
we can make concrete predictions for one-point cumulants and many-point
cumulant correlators (Munshi \& Coles 1999). We found very good
agreement between our non-linear analytical predictions and ray
tracing results for angular scales ranging from sub arcminute to tens 
of arcminutes.

Our studies also indicate that several approximations involved in 
weak lensing studies are valid even in the highly non-linear regime 
of observational interest. While
a weakly clustered dark matter distribution is expected to produce
small deflections of photon trajectories, it is not clear if the
effect of highly over-dense regions capable of producing large deflections
can also be modeled using a hierarchal ansatz
and the weak lensing approximation. Our study shows that this is indeed
the case; the effect of density inhomogeneities on very small 
angular scales can be predicted with high accuracy using
analytical approximations. Our study confirms for example that the 
Born approximation is valid in the highly nonlinear regime.

The two-point correlation functions and the projected power spectrum of 
the weak lensing convergence field $\kappa$ can be used to estimate 
the shape and normalization of the dark matter power spectrum. 
In this study we have shown 
that cosmological parameters can significantly 
affect higher order moments of the
convergence field. Earlier studies have shown that the skewness
of the convergence is sensitive to the cosmological matter density
$\Omega_m$  in the quasilinear regime (Bernardeau, van Waerbeke \&
Mellier 1997; Jain \& Seljak 1997; Schneider et al 1998), as well
as in the nonlinear regime (Jain, Seljak \& White 1999; Hui 1999). 
We have extended the work on higher order moments to include multi-point
moments and fourth order moments. The multi-point moments may be more
easily measured from non-contiguous lensing surveys.  
We find that the qualitative 
dependence of the skewness of $\kappa$ 
on $\Omega_m$  is present in these higher order statistics as well. 
In contrast, higher order moments of the 
underlying dark matter density fields are much less sensitive to background 
cosmological parameters (Jain, Seljak \& White 1999). 

The finite size of weak lensing
catalogs will play an important role in the determination of cosmological
parameters. It is therefore of interest
to incorporate such effects in future analysis. Noise 
due to the intrinsic ellipticities of lensed galaxies will also need
to be modeled. The present study is based on a 
top-hat window function which is 
easier to incorporate in analytical computations. 
It is possible to extend our study to other statistical
estimators such as $M_{\rm ap}$, which use a compensated filter to smooth
the shear field and may be more suitable for observational studies
(Reblinsky et al 1999; Schneider et al 1997). 
Our method of computing cumulants and cumulant correlators can
also be generalized to study the 
bias associated with the convergence field. 
Such studies will provide 
an interesting method of measuring
the statistics of collapsed dark objects (e.g. Jain \& Van Waerbeke 1999).
We hope to present such results elsewhere. 

It is interesting to consider an alternative to the hierarchical
ansatz for the distribution of dark matter in the highly
nonlinear regime. The dark matter can be modeled as belonging
to halos with a mass function given 
by the Press-Schechter formalism, and spatial distribution
modeled as in Mo \& White (1996).
When supplemented by a radial profile for the dark halos such a prescription
can be used to compute the one-point cumulants of the convergence field.
Reversing the argument, given the one point cumulants of the 
convergence field it is possible to estimate the statistics of dark halos. 

The one-point probability distribution
function and its two-point correlations are the complements to 
the moment hierarchy studied in this work. 
Numerical studies of the one point probability
distribution function (pdf) 
of the convergence field have been carried out
by Jain et al.(1999) and a fitting function has been provided by Yang
(1999). Recently Valageas (1999a) and Munshi \& Jain (1999) have
computed the pdf of $\kappa$ using the hierarchical ansatz and
compared the results to the measurements from simulations. 
The success of their analytical results means that the pdf for
a desired cosmological model can be computed as a function of 
smoothing angle and redshift distribution. Thus physical effects, 
such as the magnification distribution for Type Ia Supernovae, 
can be conveniently computed. Thus we have a complete analytical
description, based on models for gravitational clustering, for the
full set of statistics of interest for weak lensing -- the one point
pdf, two point correlations, and the hierarchy of higher order
cumulants and cumulant correlators. This analytical description 
has powerful applications in making predictions for a variety of
models and varying the smoothing angle and redshift distribution
of source galaxies; numerical studies are far more limited in the 
parameter space that can be explored.


\section*{Acknowledgment}
Dipak Munshi was supported by a fellowship from the Humboldt foundation at MPA
where this work was completed. 
It is a pleasure for Dipak Munshi to acknowledge 
helpful discussions with Patrick Valageas, Katrin Reblinsky, Peter Coles
and Francis Bernardeau.

\eject

\section{Appendix}

In our ray tracing simulations we have placed the source galaxies at a
redshift of unity. However it is interesting to check how these results
change when source redshift is altered. We find that the variance and
other one point cumulants are sensitively dependent on the source redshift.
While the variance increases with source redshift for a given smoothing
angle, the $S_N$ parameters and the $C_{pq}$ parameters decrease with 
the increase in  source 
redshift. For source redshifts $z_s<1$ the $S_N$ parameters are very 
sensitive to change in source redshifts. 
At low source redshifts, lensing probes the
dark matter in the highly non-linear regime whereas
for larger redshifts successive layers of matter tend to make the probability
distribution function of $\kappa$ more Gaussian, thereby reducing
the values of $S_N$ parameters. Nevertheless 
the ordering of the $S_N$ parameters 
in different cosmologies 
does not change with source redshifts. For any given source redshift
the $S_N$ parameters are largest for open models and
smallest for the TCDM model. Theoretical analysis by Valageas (1999b)
has shown that for smaller source redshifts $S_3 \propto z_s^{-1/2}$ and 
for larger source redshifts $S_3 \propto ( 1+ z_s)^{-1}$. Our
numerical results shows that this is indeed the case. Perturbative
analyses of the dependence of the $S_N$
parameters on source redshift give qualitatively similar
dependences. 

In figure 13 we plot the variance for two
different source redshifts $z_s = 0.5$ and $z_s = 1$. 
Figures 14 and 15 show $S_3$ and $S_4$ for 
redshifts $z_s = 0.5$ and $z_s = 2$; comparison with
the same quantities for $z_s = 1$ plotted in figures 3 and
4 shows that the $S_N$ parameters do decrease with source
redshift for all smoothing angles. In figures 20 and 21 we have
changed the source redshift continuously keeping the smoothing angles
fixed at $\theta_0 = 1'$ (left panels) and $\theta_0 = 2'$ (right
panels). The $S_N$ parameters show a rapid
variation for small
redshifts whereas for larger redshifts they vary at a much slower
rate. The two point analogs of the variance and $S_N$ parameters i.e.
the correlation function and $C_{pq}$ parameters shows a very similar 
pattern in their dependence on source redshifts. The two point correlation
function $\kappa(\gamma_{1})\kappa(\gamma_{2})$ is plotted in figure 16 and
the lower order cumulant correlators $C_{21}$, $C_{31}$ and $C_{22}$ in 
figures 17, 18 and 19, respectively. For small separation angles, i.e. 
small values of $\theta_{12}$, the cumulant correlators $C_{pq}$ match with
their one-point counterparts, $S_{p+q}$. 

In all of our plots we have assumed 
a specific form of the tree-level hierarchy as proposed
by Bernardeau \& Schaeffer (1992). The ansatz proposed by
Szapudi \& Szalay (1993) gives very similar results, as shown 
in figure 22. We have plotted $C_{22}$ and $C_{31}$ for $z_s
= 1$ (predictions for the third order cumulant correlator $C_{21}$ are
the same for these {\em ansatze}).

\begin{figure}
\protect\centerline{
 \epsfysize = 3.5truein
 \epsfbox[27 403 587 715]
 {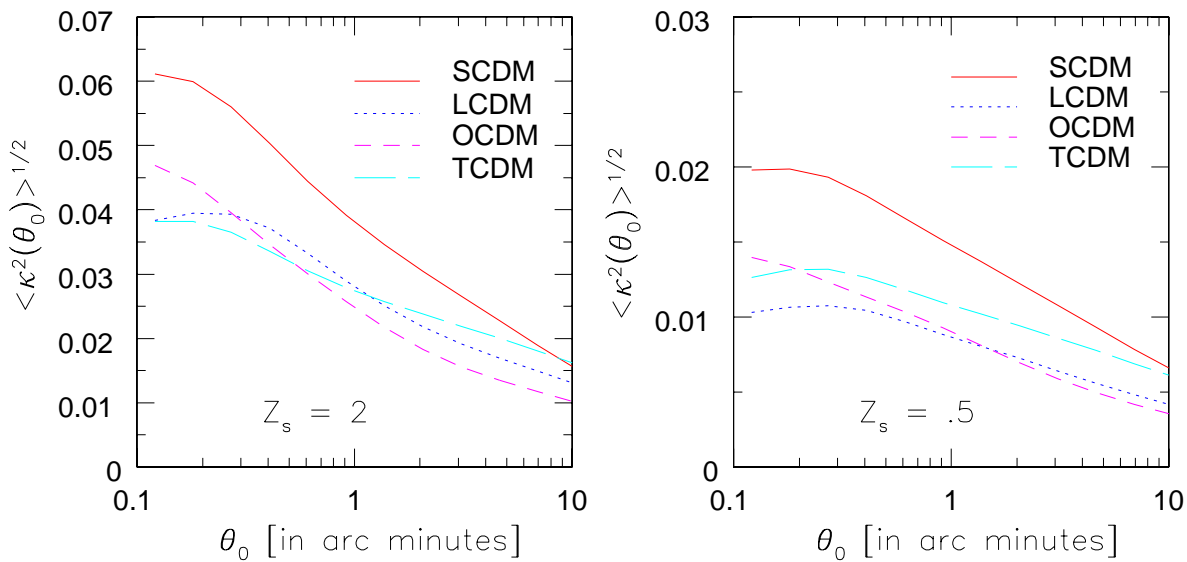} }
 \caption{The variance $\langle \kappa^2(\theta_0)\rangle$ as a 
function of smoothing angle 
$\theta_0$ for two different redshifts. Note that the variance increases 
with source redshift whereas all $S_N$ parameters decrease with 
redshift.}
\end{figure}

\begin{figure}
\protect\centerline{
 \epsfysize = 3.5truein
 \epsfbox[27 403 587 715]
 {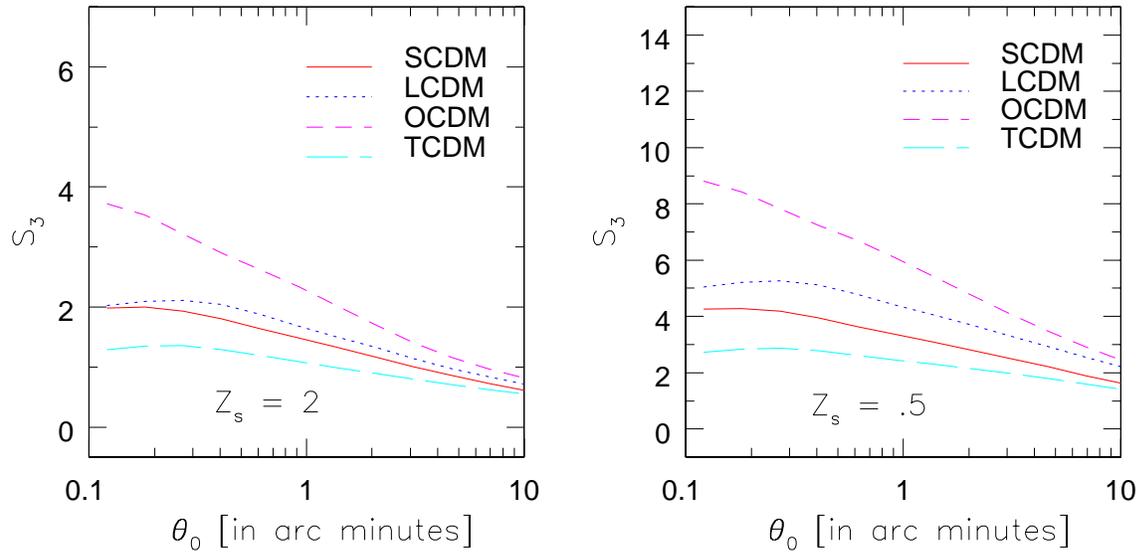} }
 \caption{The normalized skewness
$S_3$ as a function of the smoothing angle $\theta_0$
 for two different
 redshifts. $S_3$ is higher for lower source redshift
due to smaller projection effects and greater nonlinear
clustering of the lensing matter. }
\end{figure}

\begin{figure}
\protect\centerline{
 \epsfysize = 3.5truein
 \epsfbox[27 403 587 715]
 {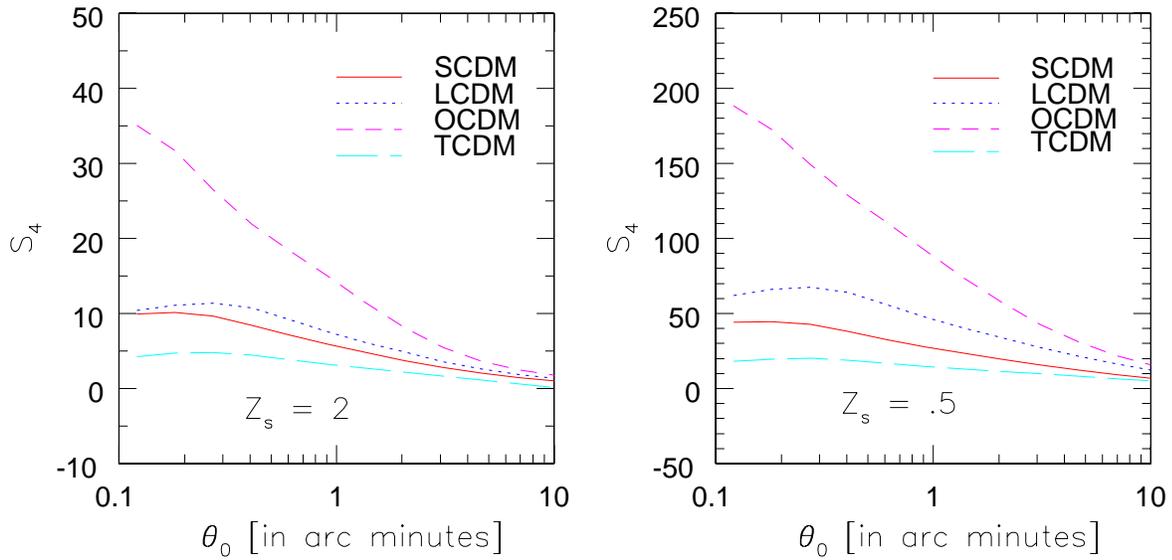} }
 \caption{The normalized kurtosis parameter $S_4$ 
 as a function of smoothing angle $\theta_0$ for two
different source redshifts. Note that the ordering of the 
different curves does not
 change with source redshift. The $S_N$ paraneters are largest for 
the OCDM model and smallest for the TCDM model for any given redshift.}
\end{figure}

\begin{figure}
\protect\centerline{
 \epsfysize = 3.5truein
 \epsfbox[27 403 587 715]
 {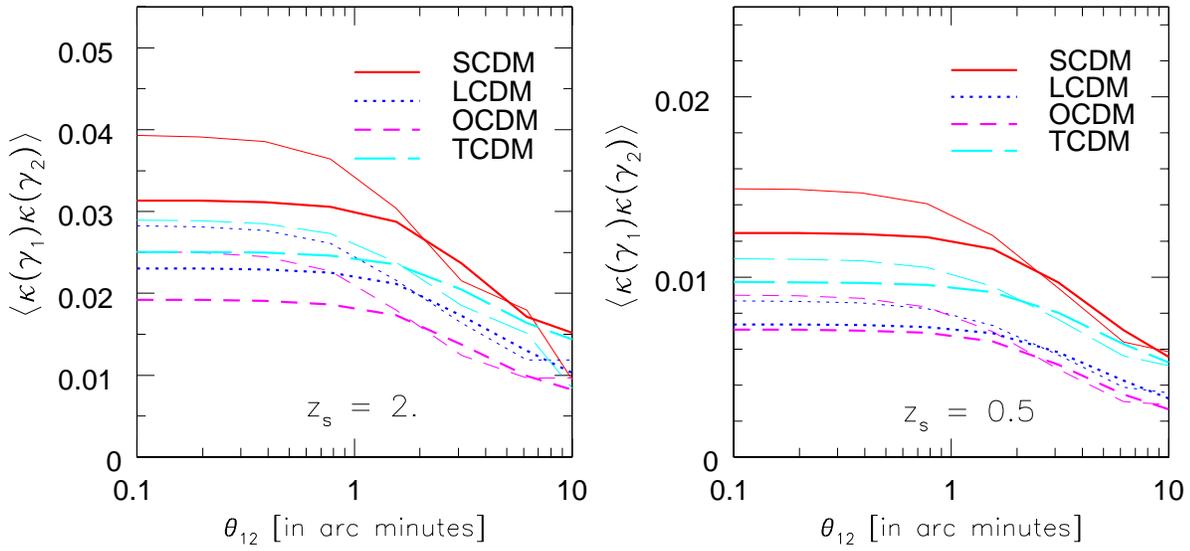} }
 \caption{The correlation function $\langle
 \kappa(\gamma_1)\kappa(\gamma_2) \rangle$  as a function of
 separation angle $\theta_{12}$ for two different source redshifts
 $z_s$. The 
smoothing angles are $\theta_0 = 1', 2'$ for the upper and lower curves at
small $\theta_{12}$, respectively. 
}
\end{figure}

\begin{figure}
\protect\centerline{
 \epsfysize = 3.5truein
 \epsfbox[27 403 587 715]
 {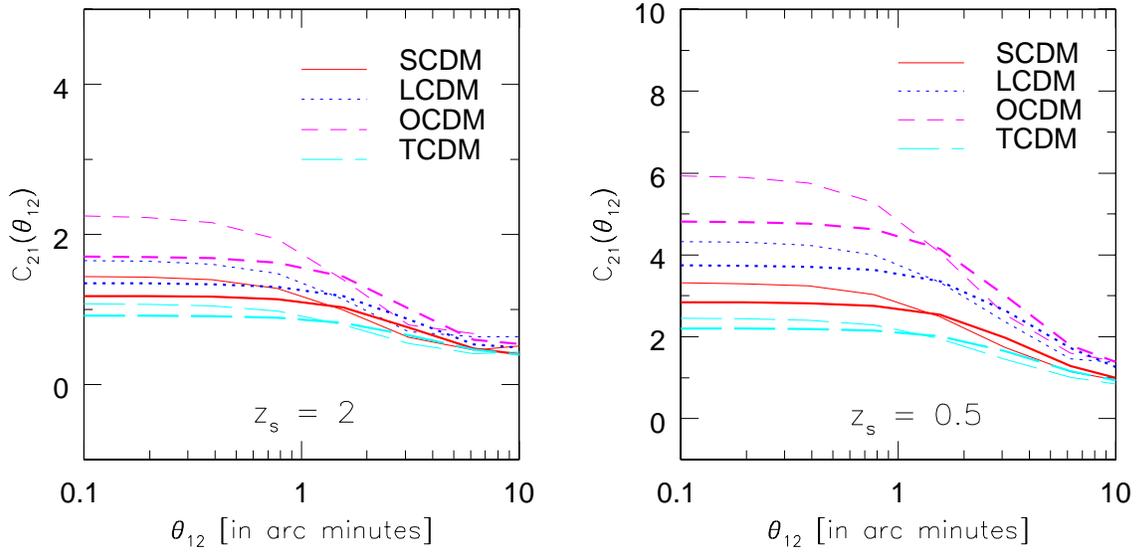} }
 \caption{The third order cumulant correlator $C_{21}$ 
as a function of separation angle $\theta_{12}$, for two different 
source redshifts $z_s$. The dependence of the cumulant correlators $C_{pq}$ on
 redshift is very similar to their one-point counterpart. }
\end{figure}

\begin{figure}
\protect\centerline{
 \epsfysize = 3.5truein
 \epsfbox[27 403 587 715]
 {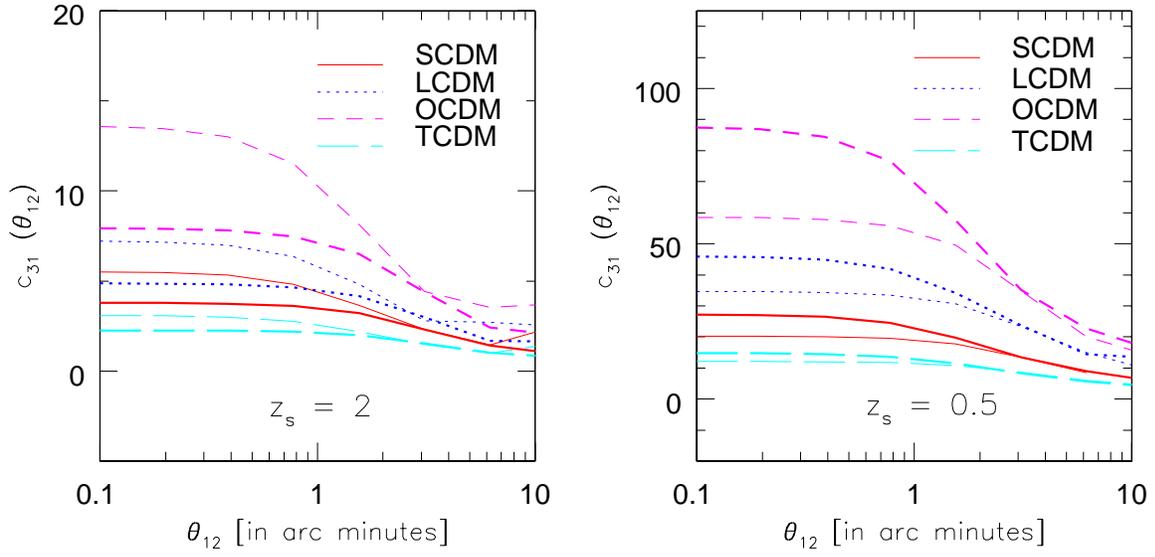} }
 \caption{The fourth order cumulant correlator $C_{31}$ for two different 
source redshifts $z_s$ as a function of separation angle
$\theta_{12}$. The 
smoothing angles are $\theta_0 = 1', 2'$ for the upper and lower curves at
small $\theta_{12}$, respectively. 
The $C_{pq}$ parameters tend to increase for low 
source redshift like the $S_N$ paramters.}
\end{figure}

\begin{figure}
\protect\centerline{
 \epsfysize = 3.5truein
 \epsfbox[27 403 587 715]
 {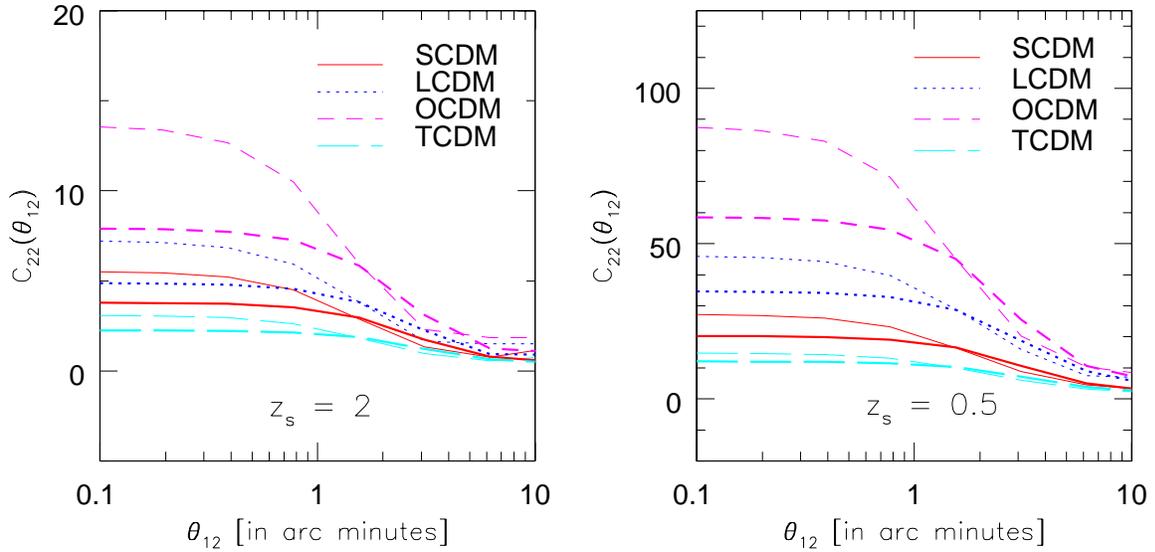} }
 \caption{The fourth order cumulant correlator $C_{22}$ for two different 
source redshifts $z_s$ as a function of separation angle
$\theta_{12}$. 
}
\end{figure}

\begin{figure}
\protect\centerline{
 \epsfysize = 3.5truein
 \epsfbox[27 403 587 715]
 {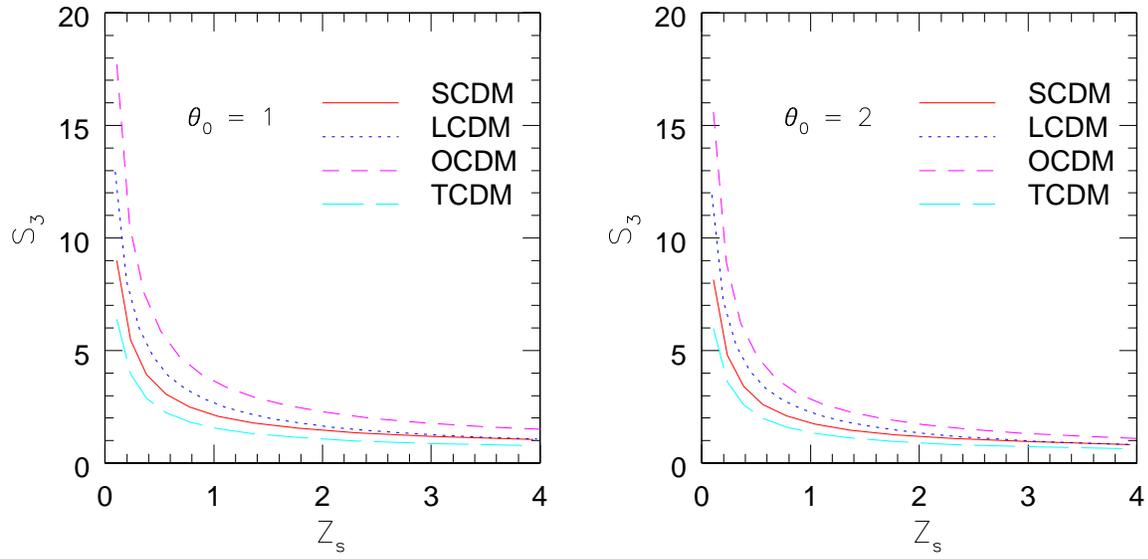} }
 \caption{Variation of $S_3$ with source redshift $z_s$ for fixed smoothing
 angle $\theta_0 = 1'$ in the left panel and $\theta_0 =
 2'$ in the right panel. While the rate of increase 
is  moderate for high redshifts the rate is 
much more rapid at smaller redshift. The growth of density inhomogeneities 
at lower redshifts causes the values of  the $S_N$ parameters to 
increase, 
while for larger redshifts subsequent layers of lensing matter
tend to make the probability distribution function more Gaussian. }
\end{figure}

\begin{figure}
\protect\centerline{
 \epsfysize = 3.5truein
 \epsfbox[27 403 587 715]
 {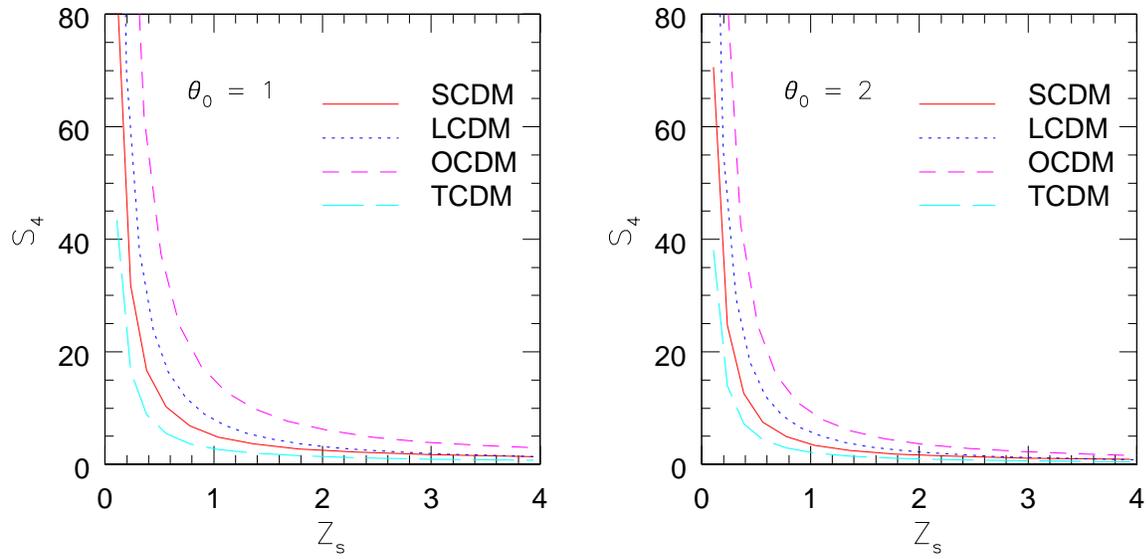} }
 \caption{ Variation of $S_4$ with source redshift
 $z_s$ for fixed smoothing
 angle $\theta_0 = 1'$ in the left panel and $\theta_0 = 2'$
in the right panel. 
}
\end{figure}

\begin{figure}
\protect\centerline{
 \epsfysize = 3.5truein
 \epsfbox[27 403 587 715]
 {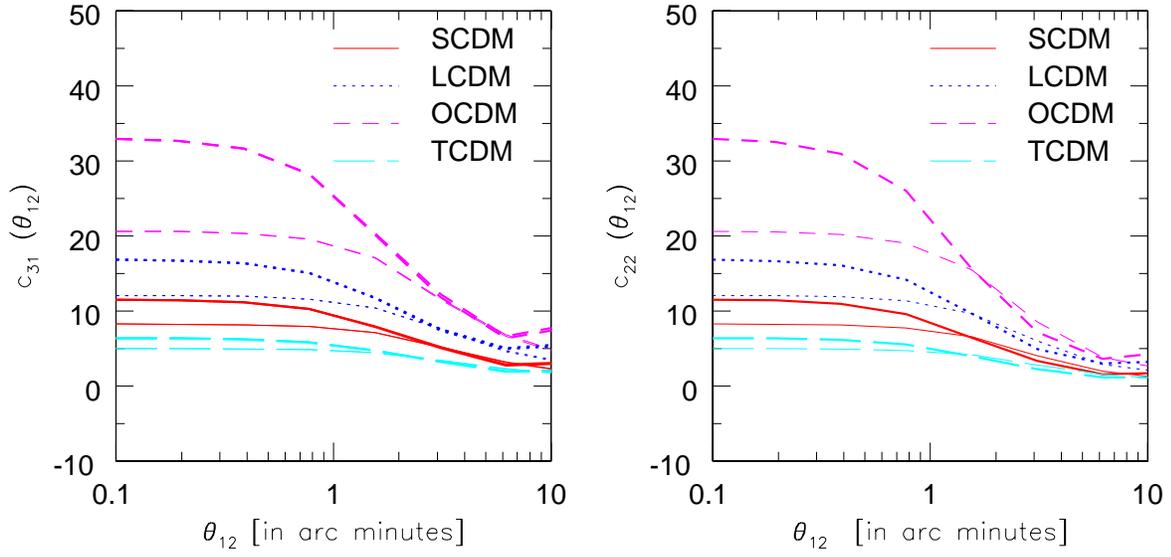} }
 \caption{The analytical predictions for the cumulant correlators
 $C_{22}$ and $C_{31}$ are plotted using the Szapudi
 \& Szalay (1993) ansatz $R_a = R_b$ in combination with hyper-extended
 perturbation theory. The 
smoothing angles are $\theta_0 = 1', 2'$ for the upper and lower curves at
small $\theta_{12}$, respectively. 
 We find that
 the predictions are almost identical to that of Bernardeau \&
 Schaeffer's ansatz, although the two differ in the way
 they assign amplitudes to trees of different topologies
 of the same order. 
}
\end{figure}

\end{document}